# Carbon Phosphide Monolayer with Superior Carrier Mobility

Gaoxue Wang[1], Ravindra Pandey[1], and Shashi P. Karna[2]

[1]Department of Physics, Michigan Technological University, Houghton, Michigan 49931, USA
[2]US Army Research Laboratory, Weapons and Materials Research Directorate, ATTN: RDRL-WM, Aberdeen Proving Ground, MD 21005-5069, U.S.A.

(January 19, 2016)

*Email: gaoxuew@mtu.edu
pandey@mtu.edu
shashi.p.karna.civ@mail.mil

**Abstract**

Two dimensional (2D) materials with a finite band gap and high carrier mobility are sought after materials from both fundamental and technological perspectives. In this paper, we present the results based on the particle swarm optimization method and density functional theory which predict three geometrically different phases of carbon phosphide (CP) monolayer consisted of $sp^2$ hybridized C atoms and $sp^3$ hybridized P atoms in hexagonal networks. Two of the phases, referred to as α-CP and β-CP with puckered and buckled surfaces, respectively are semiconducting with highly anisotropic electronic and mechanical properties. More remarkably, they have lightest electrons and holes among the known 2D semiconductors, yielding superior carrier mobility. The γ-CP has a distorted hexagonal network and exhibits a semi-metallic behavior with Dirac cones. These theoretical findings suggest the binary CP monolayer to be yet unexplored 2D materials holding great promises for applications in high-performance electronics and optoelectronics.

Since the discovery of graphene [1, 2], two dimensional (2D) materials have sparked an extraordinary level of interest due to their unique properties and novel applications in electronics and optoelectronics. Among the 2D material family, the group IV elemental monolayers, graphene, silicene and germanene stand out due to presence of the Dirac cones [3, 4], which endow the massless Dirac fermions with extremely high carrier mobility. However, the gapless nature of group IV monolayers is one of the major obstacles for their applications in transistors. Recently, the group V elemental monolayers such as phosphorene [5, 6], arsenene [7, 8] and antimonene [9, 10] were established as promising 2D materials with electronic properties which are significantly different from those of the group IV elemental monolayers. For example, phosphorene is a direct band gap semiconductor with anisotropic electronic conductance and high hole mobility [5, 11, 12]. However, due to the fast degradation of phosphorene in air, its application in electronic devices has been challenging [13-15].

Interestingly, the group IV and V elemental monolayers show noticeable structural similarities including three-fold coordinated atoms and a hexagonal network. In graphene, each C atom is $sp^2$ hybridized connecting to three neighboring C atoms in a planar hexagonal structure through $\sigma$ bonds. The out-of-plane $p_z$ orbitals form $\pi$ and $\pi^*$ bands leading to its band structure with Dirac cones [3]. In phosphorene, P atom is $sp^3$ hybridized sharing three of its valence electrons with neighboring P atoms forming a puckered hexagonal lattice. The remaining two valence electrons form a lone pair in one of the $sp^3$ orbitals. Since preference of C and P atoms appears to be three-fold coordination in the 2D monolayer, the following intriguing questions arise: Is it possible to form a stable carbon phosphide (CP) monolayer? If yes, then how will the binary monolayer be like in terms of mechanical and electronic properties including nature of the band gap and carrier mobility?

It is to be noted that experimental efforts are being made to produce carbon phosphide (or phosphorus carbide). Initial attempts to synthesize bulk CP were made by producing P-doped diamond-like carbon [16]. Later, synthesis of amorphous CP films using radio frequency plasma deposition with $CH_4$ and $PH_3$ gas mixtures was reported [17, 18]. The ratio of P/C in their samples can be widely controlled via the ratio of $PH_3/PH_4$ gas [17, 18], which led to the efforts of producing CP films using pulsed laser deposition [19, 20] and magnetron sputtering techniques [21]. In these experiments, the presence of direct C-P bonds was established. Theoretically, the properties of bulk phases of crystalline phosphide with a range of stoichiometric compositions were investigated via density functional theory [19, 22]. Various phases with three- and four-fold coordinated P atoms have been predicted [22].

To the best of our knowledge, no experimental or theoretical study has been made on CP monolayer. In this paper, we consider structure, stability, mechanical and electronic properties of the low-energy phases of CP monolayer obtained by an exhaustive structural search performed using a recently developed CALYPSO code with particle swarm optimization method [23].

The particle swarm optimization method was inspired by the choreography of a bird flock which has been demonstrated as an efficient search method for the low-energy structural phases of a material at a given composition [23]. The details of the structural searching process can be found in the supplementary materials. Figure S1(b) displays the low energy structures with different stoichiometric compositions of $C_{1-x}P_x$ monolayers with cohesive energies between those of graphene and phosphorene (Figure S1(a) in the supplementary materials). The stable monolayers are consisted of three-fold coordinated C and P atoms. The stoichiometric monolayers (x = 0.5) have attracted our particular attentions due to their compact structural configurations as shown in Figure S2(c). The stability of these monolayers is verified by the vibration spectra calculations and *ab initio* molecular dynamics (AIMD) simulations. We classify these hexagonal configurations to be α-, β-, and γ-phases of CP monolayer (Figure 1) in analogy to the classification used for phosphorene, α-P (black) and β-P (blue) monolayers [24]. The α-P has a puckered surface due to the intralayer $sp^3$ bonding character in the lattice. β-P possesses a buckled hexagonal honeycomb structure maintaining the $sp^3$ character of bonds.

Calculations of electronic properties were performed using the projector-augmented-wave (PAW) method and the generalized-gradient approximation (GGA-PBE) for electron exchange-correlation interaction [25] as implemented in the Vienna *Ab initio* Simulation Package (VASP) [26]. Since GGA usually underestimates the band gap, we also used the hybrid Heyd-Scuseria-Ernzerhof (HSE06) functional form [27] to get more accurate band gap values. The energy convergence was set to $10^{-6}$ eV and the residual force on each atom was smaller than 0.01 eV/Å. The energy cutoff for the plane-wave basis was set to 500 eV. The reciprocal space was sampled by *k*-point meshes of (11×11×1) for geometry optimization, and (45×45×1) for density of states (DOS) calculations. The vacuum distance normal to the plane was larger than 20 Å to eliminate interaction between the replicas due to the periodic boundary conditions in the supercell approach of our model. The vibration spectra calculations were performed by means of finite displacement method as implemented in the PHONOPY program [28]. The AIMD simulations were based on the NVT ensemble with a time step of 1 *fs*. The temperature was controlled to 300 K with Nośe-Hoover thermostat [29]. The structural geometry and electronic properties were also checked with calculations based on atomic-orbital bases as implemented in SIESTA [30].

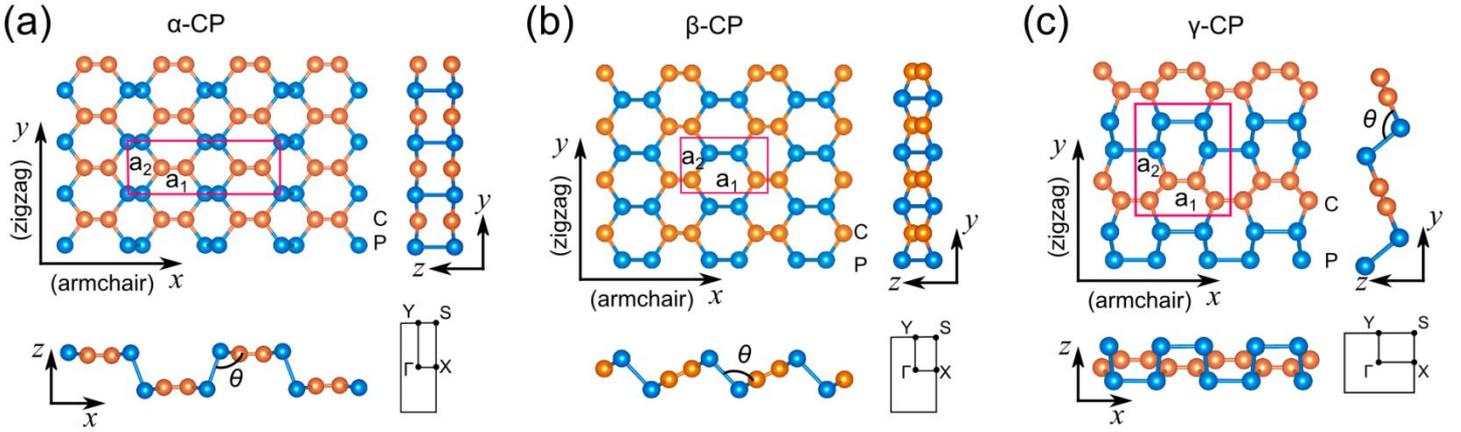

*Figure 1. The structural geometry including top view, side view, and the Brillouin zone of (a) α-CP, (b) β-CP, and (c) γ-CP. $a_1$ and $a_2$ are the lattice constants, R is the nearest neighbor distance, θ is the bond angle of C-P-P.*

*Table 1. Calculated structural parameters of CP monolayers (see Figure 1) at the GGA-PBE level of theory.*

|  | $a_1$ (Å) | $a_2$ (Å) | $R_{C-C}$ (Å) | $R_{C-P}$ (Å) | $R_{P-P}$ (Å) | θ (°) | Cohesive energy (eV/atom) |
|---|---|---|---|---|---|---|---|
| α-CP | 8.68 | 2.92 | 1.36 | 1.83 | 2.32 | 97.40 | 5.32 |
| β-CP | 4.72 | 2.91 | 1.37 | 1.82 | 2.33 | 97.78 | 5.33 |
| γ-CP | 4.80 | 5.63 | 1.45, 1.43 | 1.82 | 2.30, 2.17 | 104.00 | 5.35 |

In α-, β-, and γ-CP, each C atom bonds with three nearest neighbors in a planar configuration (see the side views in Figure 1) implying the C atoms are $sp^2$ hybridized. On the other hand, each P atom bonds with three neighboring atoms in a buckled configuration suggesting $sp^3$ hybridization of P atoms in the 2D lattice. For α- and β-CP, the zigzag (i.e. *y*) direction is composed of alternating C and P atoms, and the armchair (i.e. *x*) direction is composed of alternating C-C and P-P dimers. Overall, α-CP has a puckered surface, and β-CP has a buckled surface as seen from the side views in Figures 1(a) and 1(b). The γ-CP is composed of alternating P chain and C chain along the armchair direction (Figure 1(c)). Due to mismatch in C-C and P-P bonds, the γ-CP has a distorted hexagonal network.

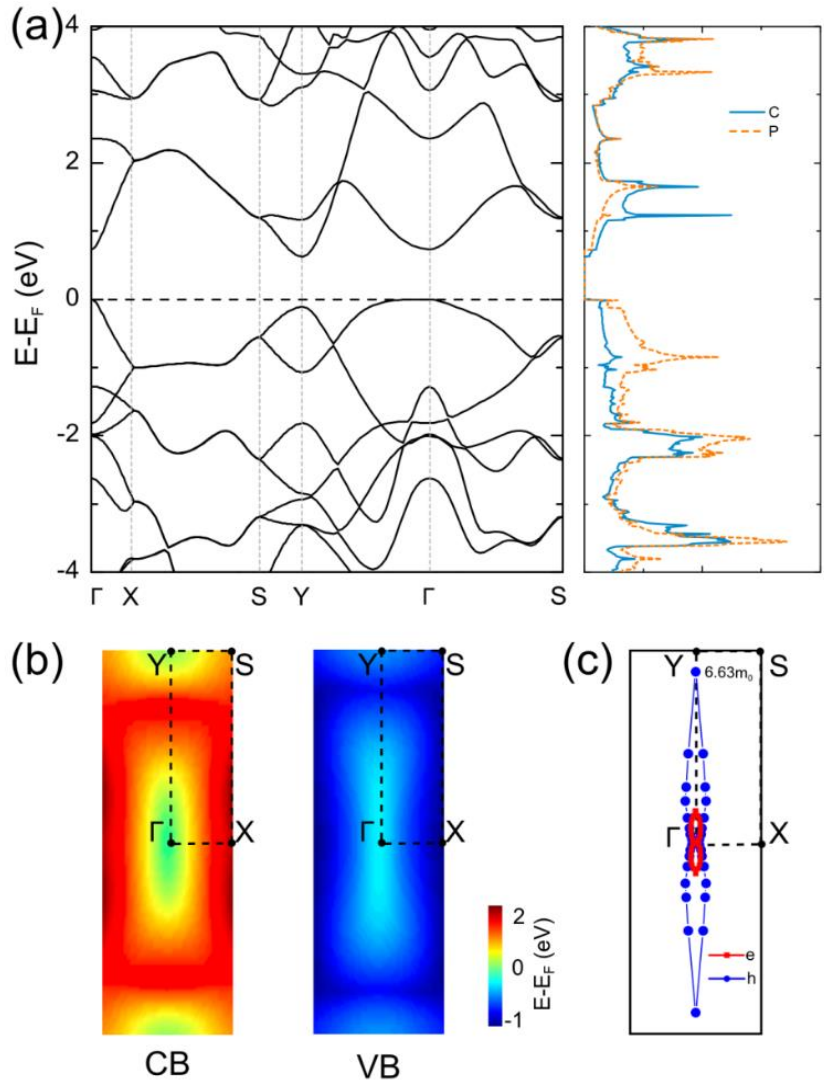

*Figure 2. Electronic properties of α-CP: (a) band structure and density of states, (b) 2D energy profiles of the first valence band (VB) and the first conduction band (CB), and (c) effective mass of electrons and holes at Γ along different directions; distance from a data point to Γ is proportional to the magnitude of the effective mass. The solid line acts as a guide to the eye.*

All three phases of CP monolayer have nearly degenerate cohesive energy with the rectangular unit cells as summarized in Table 1. The length of C-C, C-P, P-P bonds in α- and β-CP are 1.36-1.37 Å, 1.82-1.83 Å, and 2.32-2.22 Å, respectively. The C-C bond in CP monolayer is slightly shorter than that of 1.42 Å in graphene [3], and the P-P bond is slightly longer than that of 2.26 (2.22) Å in phosphorene calculated at the same level of theory [5]. In γ-CP, the length of C-C and P-P bonds vary in the range 1.43-1.45 Å and 2.17-2.30 Å, respectively which are very close to those of graphene and phosphorene. Typical C-P bond lengths of 1.85 Å were reported in the bulk CP by the GGA-PBE calculations [22].

The phonon dispersion curves (Figure S3 in the supplementary materials) show no imaginary (negative) vibration mode in the Brillouin zone; and the AIMD simulations (Videos S1, S2 and S3 in the supplementary materials) show that α-, β-, and γ-CP maintain their structural integrity up to 5 *ps* demonstrating the dynamical stability of α-, β-, and γ-CP. It is to be noticed that the slopes of the longitudinal acoustic (LA) branch along Γ-X is significantly different from those along Γ-Y near Γ (Figure S3 in the supplementary materials). The speed of sound derived from the LA branch along Γ-X (armchair) and Γ-Y (zigzag) directions are found to be (5.9, 12.0 km/s), (6.3, 12.3 km/s), and (13.3, 6.8 km/s) for α-, β-, and γ-CP, respectively, reflecting the anisotropic nature of the in-plane stiffness in the hexagonal network. The maximum vibrational frequency at 1450 cm$^{-1}$ for the optical branches of α-, and β-CP is associated with C atoms as seen in the phonon density of states affirming a high strength of C-C bonds in the hexagonal network. Additional calculations based on the strain energy curves [31, 32] (Figure S4 in the supplementary materials) reveal the in-plane stiffness along the armchair and zigzag directions to be (18.8, 171.5 N/m), (46.6, 158.3 N/m), and (233.2, 51.9 N/m) for α-, β-, and γ-CP, respectively confirming the anisotropy nature of the mechanical properties. The lower stiffness along the armchair direction is due to the puckered or buckled nature of α- and β-CP lattice (Figures 1(a) and 1(b)) which could accommodate external strains by changing the puckered or buckled angle without much distortion of the bond length. This is similar to the anisotropic mechanical properties observed for phosphorene [33]. For the γ-CP, the stiffness along the armchair direction is large because of the C chains. The in-plane stiffness of CP monolayers is smaller than that of 340 N/m in graphene [31, 34], while it is larger than that of 28.9 N/m and 101.6 N/m in phosphorene [35] (except for the α-CP along the armchair direction) due to the existence of stronger C-C and C-P bonds with $sp^2$ hybridization in the hexagonal lattice.

The electronic properties of α-CP monolayer are presented in Figure 2. The calculated band structure and density of states (DOS) indicate α-CP monolayer to be a semiconductor with an indirect band gap of 0.63 eV at the GGA-PBE level of theory. The valence band maximum (VBM) is at Γ. The conduction band minimum (CBM) is located at Y. The direct energy gap at Γ is 40 meV larger than the indirect gap from Γ to S. The C-$p_z$ and P-$p_z$ orbitals dominate the VMB and the CBM as seen in the decomposed band structure in Figure S5. The electron localization function (ELF) plots show that the electron density increases in regions around each C atom, which is the feature of the $sp^2$ hybridization of C (Figure S8(a) in the supplementary materials); the electron lone pair of each P atom is also seen in the side view of the ELF plot. Note that the hybrid HSE06 functional finds the energy gap to be 1.26 eV with the band structure (Figure S9(a) in the supplementary materials) similar to that obtained at the GGA-PBE level of theory.

The band structure and DOS indicate β-CP monolayer to be a semiconductor with a band gap of 0.39 eV (Figure 3(a)). CBM is at X point, and VBM lies very close to X point. Since the energy of the first VB at X point is only 10 meV lower than VBM, we may identify the gap to be a quasi-direct band gap for β-CP monolayer. The C-$p_z$ and P-$p_z$ orbitals dominate VBM, and C-$p_z$ and P-$p_x$ orbitals mainly contribute to CBM as seen in the decomposed band structure in Figure S6. The ELF plots (Figure S8(b) in the supplementary materials) of β-CP monolayer show the bonding characters to be similar to those in α-CP. Bader charge analyses also reveal a similar charge transfer of 0.92e and 0.99e from P atom to C atom in α- and β-CP, respectively. A larger band gap of 0.87 eV is obtained with the hybrid HSE06 functional form to density functional theory (Figure S9(b) in the supplementary materials).

α- and β-CP monolayers are found to show high anisotropy in their electronic properties. For example, valence and conduction bands around Fermi level have different slopes along the X-Γ (armchair) and X-S (zigzag) directions (Figure 3(a)), which reflects directional dependence of effective mass of electrons and holes in β-CP. From the 2D plots of the energy dispersion of the first valence and conduction bands shown in Figure 3(b), we see elongated shape along X-S. In contrast, the bands encounter a higher degree of dispersion along X-Γ reflecting a smaller effective mass of carriers.

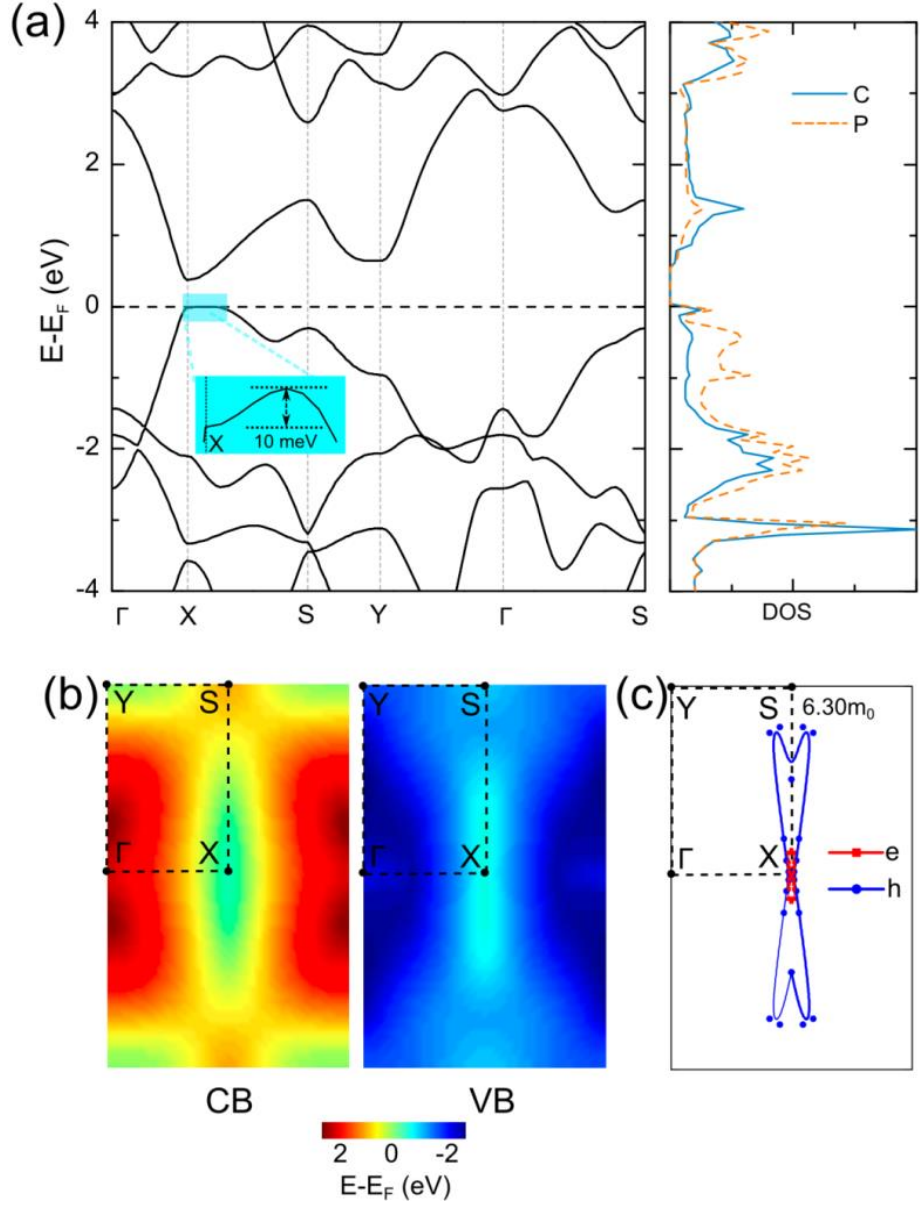

*Figure 3. Electronic properties of β-CP monolayer: (a) band structure and density of states, (b) 2D energy dispersion of the first valence band (VB) and the first conduction band (CB), and (c) effective mass of electrons and holes along different directions at X; distance from a data point to X is proportional to the magnitude of the effective mass. The solid line acts as a guide to the eye.*

The calculated directional dependence of effective mass of the carriers is shown in Figure 3(c). The values are significantly larger in the X-S (zigzag) direction than the X-Γ (armchair) direction. Along the X-Γ (armchair) direction, electrons and holes have effective masses smaller than $0.05m_0$. The values along X-S (zigzag) direction are $1.10m_0$ and $4.10m_0$, respectively. The effective mass of holes could reach a maximum of $6.30m_0$ nearby the X-S direction. α-CP also has significant anisotropic effective mass as demonstrated in Figures 2(a), 2(b) and 2(c).

The effective masses of carries in α- and β-CP along the zigzag (*y*) direction are comparable to the values in phosphorene ($1.12m_0$ and $6.35m_0$ in [12, 35, 36]), while the values along the armchair (*x*) direction are even smaller than those in phosphorene ($0.17m_0$ and $0.15m_0$ [35, 36]). The decomposed band structures in Figures S5 and S6 reveal the heavier holes along the zigzag direction (the first VB along Γ-Y and X-S directions in Figures S5 and S6, respectively) are mostly contributed by the P-$p_z$ orbital, and the lighter holes along the armchair direction (the first VB along Γ-X direction in Figures S5 and S6) are mainly associated with C-$p_z$ and P-$p_z$ orbitals. Therefore, contributions of C-$p_z$ electrons appear to decrease the effective mass of carriers in the binary carbon phosphide monolayer which are extremely important for nanoscale devices requiring semiconducting materials with high carrier mobility.

An understanding of the electronic conductance of the material can be gained from the carrier mobility calculations based on the deformation potentials (DP) theory as proposed by Bardeen and Shockley [37]. According to the DP theory, the carrier mobility of 2D materials can be evaluated according to the following expression [12, 35, 38]

$$\mu_x = \frac{eh^3 C_x}{(2\pi)^3 k_B T m_x^* m_d (E_{1x}^2)} \quad (1)$$

where *e* is the electron charge, *h* is the Planck's constant, *T* is the temperature and $m^*$ is the effective mass. $m_d$ is determined by $m_d = (m_x^* m_y^*)^{1/2}$. $E_{1x}$ is the deformation potential defined as $E_{1x} = \Delta V/(\Delta a_x/a_x)$, and is obtained by varying the lattice constant $a_x$ along the direction of electron conduction. $\Delta V$ is the change of the band energy. The in-plane stiffness constant $C_x$ is obtained by evaluating the strain energy curve [31, 32]. Equation 1 has been demonstrated previously to give a reliable estimate for the upper limit of the carrier mobility in phosphorene

[12, 35].

Table 2. Calculated carrier mobility in α-CP monolayer at T = 300 K along x (armchair) and y (zigzag) direction. $m_e^*$ and $m_h^*$ are the effective masses of electron (e) and hole (h), respectively.

|   | $m_e^*/m_0$ | $m_e^*/m_0$ | $E_{1x}$ | $E_{1y}$ | $C_x$ | $C_y$ | $\mu_x$ | $\mu_y$ |
|---|---|---|---|---|---|---|---|---|
|   | x | y | (eV) | | (Nm$^{-1}$) | | ($10^3$ cm$^2$V$^{-1}$s$^{-1}$) | |
| e | 0.10 | 1.22 | 1.72 | 10.55 | 18.75 | 171.47 | 3.87 | 0.08 |
| h | 0.12 | 6.63 | 0.18 | 1.75 | 18.75 | 171.47 | 115.18 | 0.20 |

Table 3. Calculated carrier mobility in β-CP monolayer at T = 300 K along x (armchair) and y (zigzag) direction. $m_e^*$ and $m_h^*$ are the effective masses of electrons (e) and holes (h), respectively.

|   | $m_e^*/m_0$ | $m_e^*/m_0$ | $E_{1x}$ | $E_{1y}$ | $C_x$ | $C_y$ | $\mu_x$ | $\mu_y$ |
|---|---|---|---|---|---|---|---|---|
|   | x | y | (eV) | | (Nm$^{-1}$) | | ($10^3$ cm$^2$V$^{-1}$s$^{-1}$) | |
| e | 0.05 | 1.10 | 2.56 | 9.30 | 46.56 | 158.27 | 12.91 | 0.15 |
| h | 0.05 | 4.10 | 1.68 | 1.66 | 46.56 | 158.27 | 15.52 | 0.66 |

The calculated carrier mobility using equation 1 at room temperature (T = 300K) for α- and β-CP is summarized in Tables 2 and 3, respectively. The carrier mobility shows strongly directional dependence as one would expect from the anisotropic nature of the calculated effective masses for α- and β-CP. The electron and hole mobility along the armchair direction is significantly larger than those obtained along the zigzag direction suggesting the presence of an anisotropic conductance in α- and β-CP. Such strong anisotropy in carrier mobility can be measured in experiments [5] and may facilitate fabrication of the anisotropic electronic devices. More interesting than the anisotropic electronic conductance is the large value of the carrier mobility in CP monolayers. For example, the hole mobility at room temperature in α-CP could potentially reach $1.15 \times 10^5$ cm$^2$V$^{-1}$s$^{-1}$, which is approximately five times larger than the maximum value in phosphorene ($0.26 \times 10^5$ cm$^2$V$^{-1}$s$^{-1}$ [35]), and significantly larger than other 2D materials, such as MoS$_2$ [39]. Such a large hole mobility in α-CP is attributed to a small effective mass together with a small deformation potential along the armchair direction (Table 2). β-CP has comparable hole mobility to phosphorene, while the electron mobility is much larger that in phosphorene [35].

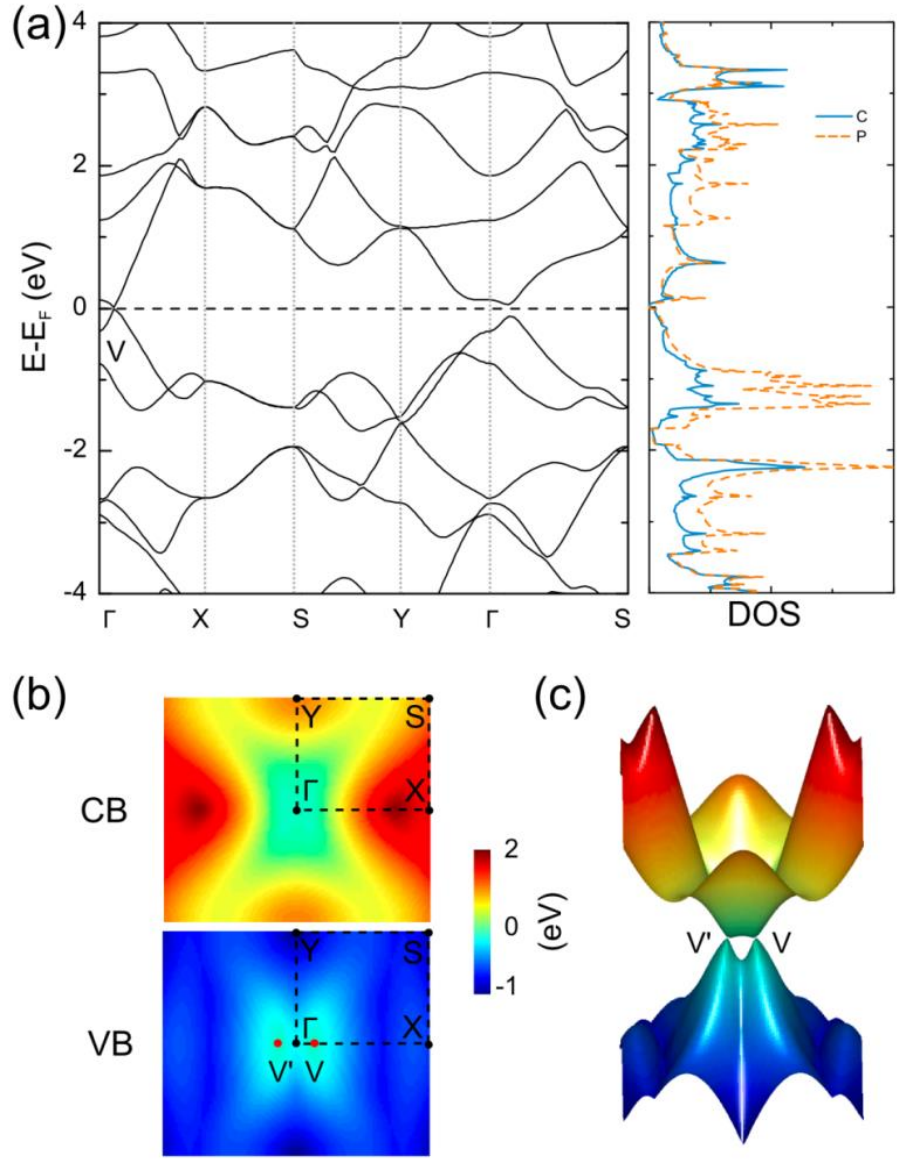

*Figure 4. Electronic properties of γ-CP monolayer: (a) band structure and density of states, (b) 2D energy dispersion of the first CB and first VB, and (c) 3D plot for first VB and first CB. V and V' are the Dirac points in the Brillouin zone.*

Distinct from α- and β-CP, γ-CP is found to be a semimetal as shown in Figure 4. From the band structure in Figure 4(a), the VB and CB cross at V point on the Γ-X. The 2D and 3D energy dispersion plots in Figures 4(b) and 4(c) illustrate that VB and CB touch at V and V' in the Brillouin zone forming distorted Dirac cones. An average of 0.46e is transferred from P atom to C atom from Bader charge analysis. The decomposed band structure in Figure S7 demonstrates that overlapping of C-$p_z$ and P-$p_z$ orbitals is responsible for the emergence of Dirac cones. The calculated Fermi velocity ($v_k = \left(\frac{1}{\hbar}\right)\left(\frac{\partial E_k}{\partial k}\right)|_{E_k=E_F}$) of electrons and holes along the V-X direction is 0.80 ×10$^6$ m/s and 0.45×10$^6$ m/s, respectively. The Fermi velocity of electrons is close to the value in graphene ($v_F$ =0.85×10$^6$ m/s [40]) implying the high electron mobility in γ-CP. Calculations based on the hybrid HSE06 functional further verify the formation of the Dirac cones in γ-CP (Figure 9S(c) in the supplementary materials). Therefore, γ-CP is like graphyne showing Dirac Fermions in a rectangular lattice [31, 41].

The most appealing properties which make carbon phosphide monolayers intriguing members of the 2D material family are the anisotropic nature of electronic conductance and high carrier mobility. α- and β-CP monolayers are predicted to have strongly anisotropic electronic properties together with the smallest effective mass of carriers (≈0.10$m_0$, and 0.05$m_0$, respectively) among the known 2D semiconductors such as phosphorene (0.15$m_0$) [35, 36] and MoS$_2$ (0.45$m_0$) [42].

α-, β-, and γ-CP monolayers cannot be fabricated with the mechanical exfoliation methods due to the absence of layered bulk counterparts. The possible synthesis approach can be chemical vapor deposition (CVD) which has been successfully used to synthesize 2D materials including group IV monolayers, such as graphene [43, 44] and silicene [45, 46], and group V thin films, such as Bi(111) and Bi(110) [47, 48]. We also notice recent progress in the fabrication of carbon phosphide thin films including strong evidence of the formation of C-P bonded regions in samples prepared with the pulsed laser deposition method has been found [20]; The black phosphorus-graphite composites with *sp$^2$* hybridized P-C bonds using mechanical milling process have also been reported [49]. The present work will further inspire the experimental realization of CP monolayers.

In summary, structure, stability and electronic properties of CP monolayers, namely α-, β-, and γ-CP, have been predicted. The structural configurations are comprised of *sp$^2$* hybridized C and *sp$^3$* hybridized P atoms in hexagonal networks. α- and β-CP are semiconductors with high anisotropy in electronic and mechanical properties. A large carrier mobility is predicted due to the small effective mass of the carriers. γ-CP is semi-metallic with Dirac cones. Our results suggest the yet unexplored binary carbon phosphide monolayers to hold great promises for applications in high-performance electronics and optoelectronics at nanoscale.

**ASSOCIATED CONTENT**

**Supporting Information Available:** Figures S1 Cohesive energy and structure of CP monolayer with various stoichiometric compositions obtained using CALYPSO, Figures S2 History of CALYPSO steps and structure of CP monolayer, Figures S3 Phonon dispersion, Figures S4 Strain energy curves, Figures S5 Projected band structure for α-CP, Figure S6 Projected band structure for β-CP, Figure S7 Projected band structure for γ-CP, Figure S8 ELF plots, Figure S9 Band structures obtained with GGA-PBE and HSE06 functional; Video S1 AIMD simulation of α-CP at 300 K, Video S2 AIMD simulation of β-CP at 300 K, Video S3 AIMD simulation of γ-CP at 300 K.

**ACKNOWLEDGEMENTS**

RAMA and Superior, high performance computing clusters at Michigan Technological University, were used in obtaining results presented in this paper. Supports from Dr. S. Gowtham are gratefully acknowledged. This research was partially supported by the Army Research Office through grant number W911NF-14-2-0088.

Table of Contents (TOC) Graphics

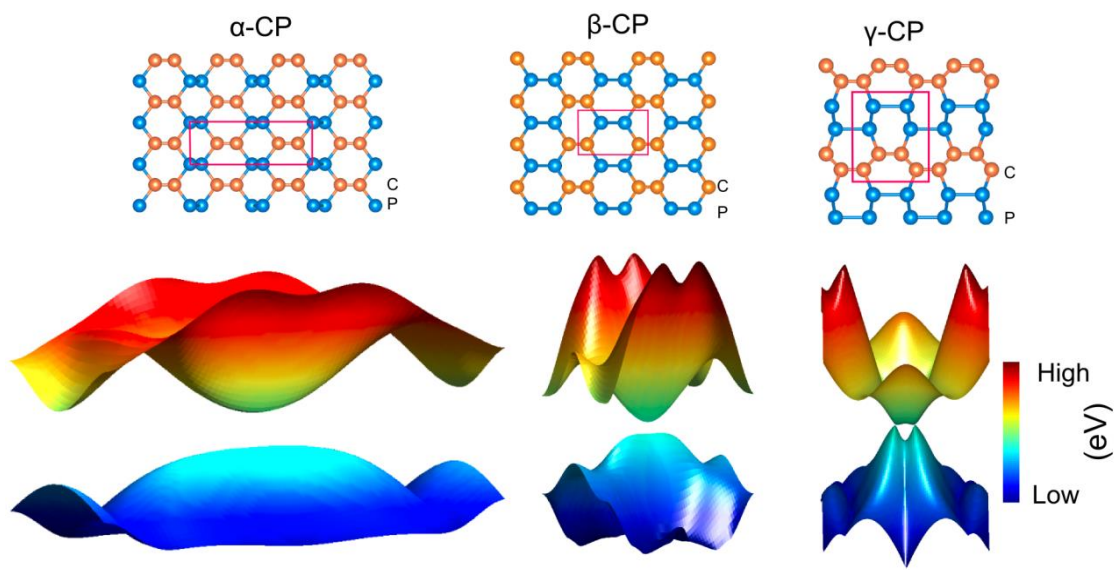